\documentclass[%
reprint,
jmp,%
amsmath,amssymb,
preprint,%
reprint,graphicx
]{revtex4-1}

\draft 

\usepackage{graphicx}
\usepackage{dcolumn}
\usepackage{bm}%
\usepackage{epstopdf}

\begin{document}
	
	\preprint{TSU}
	
	\title{Transition form-factor of $\gamma\rightarrow 3\pi$ in nonlocal quark model.} 
	
	
	
	\author{A.~S.~Zhevlakov}
	\email{zhevlakov@phys.tsu.ru}
	\affiliation{Department of Physics, Tomsk State University, 634050 Tomsk, Russia}
	
	
	\date{\today}
	
	\begin{abstract}
		The transition form factor of a gamma into three pions  $F_{3\pi}$ is studied in a framework of nonlocal chiral quark model. In local limit the result is in agreement with chiral perturbative theory and reproduce   Wess-Zumino-Witten anomaly. Nonlocality does not change a value of transition form factor in chiral limit. 
		On the physical threshold of reaction the form factor $F_{3\pi}$ obtained is in a good agreement with experimental data if the current quark mass is taken into account in model.
		
	\end{abstract}
	
	
	\maketitle 
	
	\section{\label{sec:level1} Introduction}
	
	Starting from a paper of J.~Steinberger in 1949 \cite{Steinberger:1949wx} anomalies have played an important role in investigation of strong-interaction physics at  low-energies. In case of chiral model, anomalous Ward identities are due to Wess-Zumino-Witten (WZW) effective action \cite{Wess:1971yu}. WZW action  can be constructed in terms of pseudoscalar fields - Goldstone bosons and interaction vertices involving an odd number of Goldstone bosons. Using  a topology arguments,  E.~Witten showed \cite{Witten:1983tw} that this action can be quantized and i.e. a multiple of an integer parameter $n$ which associated also with baryon number in soliton models. 

J.~Wess and B.~Zumino considered the hadronic system in presence of external vector and axial vector fields\cite{Wess:1971yu}. As a result the anomalous Ward identities and vertex of interaction of Goldstone bosons with external vector field were  obtained. All amplitudes that follow from of WZW action are given entirely in terms of the electric charge $e$ and the weak pion decay constant $f_\pi$.
 	
A well-studied process that follows from WZW action is decay of a neutral pseudoscalar meson into two photons or production of a neutral pseudoscalar (PS) mesons in $e^+e^-$ collisions. This transition form-factor was studied using a number of different approaches: chiral perturbation theory (ChPT), vector meson dominance approach \cite{Arriola:2010aq,Knecht:2001qf}, sum rules \cite{Mikhailov:2010ud, Oganesian:2015ucv}, different local and nonlocal quark models \cite{Dorokhov:2011zf, Dumm:2013zoa}.   
According to WZW action, the amplitude of this reaction has the following form
\begin{eqnarray}
A(\pi^0\rightarrow\gamma\gamma)
=F_{\gamma\gamma}(M^2_{\pi^0})\,\epsilon^{\mu\nu\alpha\beta}\epsilon^{\mu} k_1^{\nu} \epsilon^{\alpha}k_2^{\beta},
\end{eqnarray}
where $\epsilon^{i}_j$ and $k^i_j$ - polarizations and momenta of photons, transition pion-photons form factor 
\begin{eqnarray}
F_{\gamma\gamma}(0)=\dfrac{e}{4\pi^2f_\pi}.
\end{eqnarray}
The weak pion decay constant is $f_\pi=f_0[1+O(m_q)]=92.4 \, \mathrm{MeV}$. The quark mass correction in ChPT is a small \cite{Bernstein:2011bx} but plays an important role in Dalitz decay $\pi^0\rightarrow \gamma e^+e^-$ As it was shown in reference \cite{Bijnens:1991db}. On a mass-shell form-factor  $F_{\gamma\gamma}$  have a corrections due to mass of pion.
 
Another processes which are also given entirely in terms of the electric charge $e$
and the pion decay constant $f_\pi$ is the reaction of $\gamma\pi^\pm \rightarrow \pi^\pm\pi^0$ or $e^+e^-\rightarrow\gamma^*\rightarrow \pi^0\pi^-\pi^+$. 
These processes has also a connection to the WZW anomalous effective action and amplitude of reaction have the same following form:
\begin{eqnarray}
A(\gamma\pi^-\rightarrow \pi^-\pi^0)=-iF_{3\pi}(s,t,u)\,\epsilon^{\mu\nu\alpha\beta}\epsilon^{\mu} p_0^{\nu} p_1^{\alpha}p_2^{\beta},
\end{eqnarray}
where $\epsilon^\mu$ is polarization of incident photon and $p_i$ are momenta of pions. In chiral limit in low-order by quark-loops, form-factor of this amplitude is independent from  Mandelstam variables $s,t,u$ and have a simple form \cite{Giller:2005uy}
\begin{eqnarray}
\label{low_energy_theorem}
F_{3\pi}(0,0,0)=\dfrac{e}{4\pi^2f_\pi^3}=9.72 \,\mathrm{GeV}^{-3}.
\end{eqnarray}
The process that described this amplitude was measured for the first time at the IHEP accelerator (Serpukhov) at the 40-GeV
negative-pion beam \cite{Antipov:1986tp}. The experiment was based on pion pair production by pions in the nuclear
Coulomb field via the Primakoff reaction
\begin{eqnarray}
\label{Primakoff}
\pi^- \, + \, (Z,A)\, \rightarrow \, \pi^{-\prime}  + \, (Z,A)\, + \, \pi^0.
\end{eqnarray}
The value of $F_{3\pi}$ extracted from the experiment is  
\begin{eqnarray}
F_{3\pi}^{\mathrm{exp}}=(12.9\pm 0.9\pm 0.5)\, \mathrm{GeV}^{-3}.
\end{eqnarray}
This expression still has a small difference from the theoretical value given by low-energy theorem(\ref{low_energy_theorem}). The one-photon exchange is a dominating process in this reaction. One can expect that the higher precision in measuring of this value will be obtained from the future experiments. Therefore, there is a need in more accurate theoretical estimations of this value.



From the experiment side of reaction (\ref{Primakoff}) will be measured in CERN COMPASS experiment \cite{CERN_Baum} for scattering kaons and pions on nuclear target. 
There exist theoretical predictions on the value of the formfactor (5) in the framework of different approaches, e.g., low-energy theorem, ChPT (with and without taking to account the $q^2$-dependence and electromagnetic corrections)\cite{Ametller:2001yk}, dispersion relations and others.

The present work will be focus of the transition form-factor for $\gamma^*\rightarrow\pi^+\pi^0\pi^-$ which will be described in the framework of a nonlocal chiral quark model (N$\chi$QM). This model is a nonlocal extension of the Nambu-Jona Lasinio (NJL) model \cite{Anikin:2000rq,Dorokhov:2000gu,Noguera:2005ej,Noguera:2008cm,Scarpettini:2003fj}. The N$\chi$QM model has tested in different regions of particle physics: meson dynamics \cite{Dorokhov:2011zf,Noguera:2008cm}, deconfinement \cite{Radzhabov:2003hy}, dense matter at high temperature and/or intense magnetic fields \cite{Radzhabov:2010dd, GomezDumm:2005hy}.    

 In this work, the use of this model results in reproduction of low-energy theorem for the process photon into to three pions. Due to nonlocality of this model the result obtained is in a good agreement with the experimental data from Serpuhov \cite{Antipov:1986tp}.

This paper is organized as follows. In sect. \ref{sec:level2}  the nonlocal quark model is briefly considered.  In sect. \ref{sec:level3} the calculation of transition form-factor of photon into tree pions decay are performed in the framework of this model and some properties of results obtained are discussed.
In sect. \ref{sec:Outlook} the summary is given and some conclusions are made. Technical details can be found in appendix.

 	\section{\label{sec:level2}N$\chi$QM}

The Lagrangian of the $SU(2)\times SU(2)$ nonlocal chiral quark model has the
form \cite{Anikin:2000rq, Scarpettini:2003fj}
\begin{eqnarray}
	\mathcal{L}_{N\chi QM} &&=\bar{q}(x)(i\hat{\partial}-m_{c})q(x)+\\
&&	\dfrac{G}{2}[J_{S}%
	^{a}(x)J_{S}^{a}(x)+J_{P}^{a}(x)J_{P}^{a}(x)]. \nonumber
\end{eqnarray}	
where $q\left(x\right)$ are the quark fields, $m_{c}$ is the diagonal
matrix of\ the quark current masses \footnote{We consider the isospin limit
	where masses of quarks $u$ and $d$ are equal.} and $G$ is the four-quark
coupling constant.

The nonlocal structure of the model is introduced via the
nonlocal quark currents
\begin{eqnarray}
J_{S,P}^{a}(x)=&&\int d^{4}x_{1}d^{4}x_{2}\,f(x_{1})f(x_{2})\,\bar{q}%
(x-x_{1})\, \nonumber
\Gamma_{S,P}^{a}q(x+x_{2}),\\
&&\quad\Gamma_{{S}}^{a}=\tau^{a}%
,\Gamma_{{P}}=i\gamma^{5}\tau^{a},
\end{eqnarray}
where $f(x)$ is a form factor reflecting the nonlocal properties of the QCD vacuum and $\tau^{a}$ are Pauli matrices. For simplicity in this work we do not consider an extended model \cite{Villafane:2016ukb,Noguera:2008cm}
that includes other structures besides the pseudoscalar (P) and scalar (S) ones.

The form factor $f(x)$ characterizing the composite structure of hadron is an unknown function. The choice of this form factor can be sensitive for physical observables \cite{Anikin:1993mm}. To simplify the calculations the Gaussian \cite{Scarpettini:2003fj, Anikin:1993mm} form of the form factor function will be used. This function describe a nonlocality of interaction of quark field with mesons.
\begin{eqnarray}
f(x)=\int \frac{d^4 k}{(2\pi)^4}e^{ixk}f(k^2), \qquad  f(k^2)= e^{k^2/\Lambda^2},
\end{eqnarray}    
where $\Lambda$ is a parameter cutoff. Note that this form is not the only one that can be used for parametrization quark-antiquark
potential \cite{Anikin:1993mm, Noguera:2008cm}.

Observable degrees of freedom are meson states. Procedure of introducing of meson states is due to using properties of partition function $Z=\int Dq D\bar{q}\exp[-S_E]$ and trick of Gaussian integral. Integrating out the quark fields one can see that produced functional have a form: 
\begin{equation}
Z=\int D\vec{\pi} D\sigma \, \exp[-S^{(\sigma,\pi)}_E],
\end{equation}
where bosonisated action is
\begin{equation}
S^{(\sigma,\pi)}_E = -\ln\det(\textbf{\textit{D}})+\frac{1}{2G}\int \frac{d^4 p}{(2\pi)^4} (\sigma^2 +\vec{\pi}^2).
\end{equation}
The operator $\textbf{\textit{D}}$ in momentum space can be written as
\begin{eqnarray}
\textbf{\textit{D}}=(-\hat{p}-m_c)(2\pi)^4\delta(p-p^\prime) +f(p^2)f(p^{\prime 2})(\sigma+\tau^a\pi^a),\nonumber
\end{eqnarray}
where $f(p)$ is Fourier transform of the form factor $f(x)$. Assuming that scalar field $\sigma$ has a non-trivial vacuum average, while the mean field (MF) values of the
pseudoscalar fields $\pi^a$ are zero, it is possible rewritten in terms of new scalar and pseudoscalar fields as follows:
\begin{eqnarray}
\sigma(x) = m_d+\sigma(x); \qquad \pi^a(x)=\pi^a(x).
\end{eqnarray}
This proposal of non-trivial vacuum average of scalar field spontaneously breaks of symmetry. 
Parameter  $m_d$ is dynamical mass of quark. 
%

Bosonized effective action can be expanded in powers of the meson fluctuations and the expression takes the form 
\begin{equation}
S^{(\sigma,\pi)}_E = S_E^{MF}+S_E^{quad}+...
\end{equation}
For calculation of considered transition form-factor one needs to concentrate on mean field action per unit volume and action in the quadratic term for the pseudoscalar fields. These are given by 
\begin{eqnarray}
&&\frac{S^{MF}_E}{V^{(4)}}=-4N_c \int \frac{d^4 p}{(2\pi)^4}\ln[p^2+m^2(p^2)]+\frac{m_d^2}{2G},\\
&& S_E^{quad}= \frac{1}{2} \int \frac{d^4 p}{(2\pi)^4} G^-(p^2)\vec{\pi}(p)\cdot\vec{\pi}(-p). \nonumber
\end{eqnarray}	
New notation are introduced here 
\begin{eqnarray}
	&& m(p^2)=m_c+m_df^2(p^2), \label{mass_q}\\
	&& G^-(p^2)=\frac{1}{G}-8N_c\Pi_a(p^2),
\end{eqnarray}
where the mass of quark has in dependence on momentum. The  polarization operator  $\Pi_a(p^2)$ is
\begin{equation}
\Pi_a(p^2)=\int\frac{d_{E}^{4}k}{(2\pi)^{4}}\frac{f^{2}_{k_{+}}f^{2}_{k_{-}}\left[  (k_{+}\cdot k_{-})\pm m(k_{+}^{2})m%
	(k_{-}^{2})\right]  }{\left[  k_{+}^{2}+m^{2}(k_{+}^{2})\right]  \left[
	k_{-}^{2}+m^{2}(k_{-}^{2})\right]  },
\end{equation} 
with  $k_{\pm}=k\pm p/2$ and $f_{k_i}=f(k_i^2)$, index $a$ corresponds to different channels of meson. The upper
sign in numerator corresponds to the pseudoscalar channel and the lower sign
corresponds to the scalar channel. The integration here and later is gone on Euclid space $d^4_E k$ and $k^2=-k_E^2$.

\subsection{Fitting the model}

After introducing the dynamical mass of quarks, the model have four parameters: current $m_c$ and dynamical $m_d$ quark masses,  four-quark interaction  constant $G$ and  cutoff parameter $\Lambda$. These parameters are not independent. There have connections which can be obtained as follows. Using the properties of action $S$ for composite operators, one can construct an equation for the size parameter (so-called gap equation):
\begin{equation}
\left\langle\frac{\partial S}{\partial \sigma}\right\rangle_0=0.
\end{equation} 
The latter relation leads to the following expression for dynamic quark mass:
\begin{equation}
m_d=8N_c G \int\frac{d_{E}^{4}p}{(2\pi)^{4}} \frac{f^2(p^2)m(p^2)}{p^2+m^2(p^2)}.
\end{equation}	   
One parameter can remain free, for example, dynamical mass. This parameter can vary in its appropriate physical range. Two parameters: current $m_c$ and $G$ or $\Lambda$can be fitted using pion mass and width of pion into two photons decay. Last parameter can  be found from the gap equation.
 
The chiral condensates are given by the vacuum expectation values $\langle\bar{q}q \rangle = \langle\bar{u}u \rangle = \langle\bar{d}d \rangle$.
By performing  the variation of mean field partition function with respect to
the corresponding current quark masses is obtained that
\begin{equation}
\langle\bar{q}q \rangle=-4N_c\int\frac{d_{E}^{4}p}{(2\pi)^{4}} \left(\frac{m(p^2)}{D(p)} -\frac{m_c}{p^2+m_c^2}\right),
\end{equation}
where $D(p)=p^2+m^2(p^2)$.

\begin{figure}[th!]
	\includegraphics[ scale=0.34]{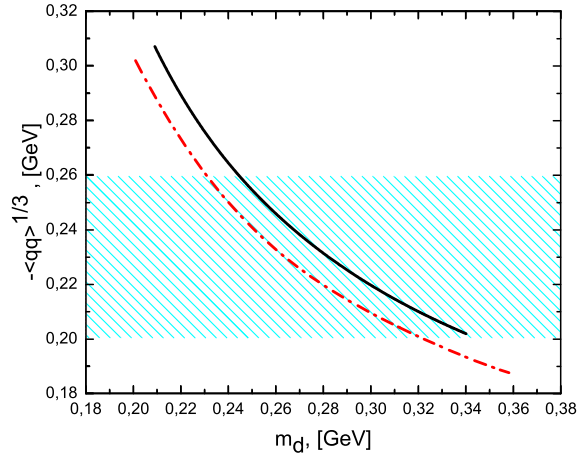}%
	\caption{\label{qq_cond} Behavior of quark condensate $-\langle\bar{q}q\rangle^{1/3}$ from dynamical mass of quark in N$\chi$QM model. Black solid line corresponds a chiral limit. Red dot-dashed line shows a dependence at physical current quark mass. Dashed box corresponds the emperical bounds from \cite{Dosch:1997wb,Giusti:1998wy}. }
\end{figure}

Value of chiral condensate of light quarks is one of the fundamental parameters of non-perturbative QCD
and chiral symmetry. Detailed analysis of quark condensate behavior in nonlocal quark model with different parametrizations was provided in \cite{GomezDumm:2006vz}. Typical interval for physical dynamical mass of quark in the N$\chi$QM model is taken in range $200 \div 350$ MeV  as reproduces empirical bands for light quark condensate $-\langle\bar{q}q\rangle^{1/3} \simeq 200 \div 260 $ MeV \cite{Shifman:1978bx,Dosch:1997wb,Giusti:1998wy}. In addition physical range in N$\chi$QM model with Gaussian function of description a nonlocality of quark interaction with meson has qualitative agreement with  the results  of quark condensate  from lattice calculation \cite{Aoki:2013ldr, Colangelo:2010et} and result from renormalization group optimized perturbation method \cite{Kneur:2015dda}.  Behavior of light quark condensate from dynamical quark mass $m_d$ in  N$\chi$QM model is shown in Fig.\ref{qq_cond}. Behavior of quark condensate at physical value of current quark mass has same view of curve as one in chiral limit but lies just below. 

The error bar of  estimation different values in framework of N$\chi$QM model comes from the band in region of physical dynamical mass of quark.

\subsection{Nonlocal vertices}

After we apply of mechanism of spontaneous symmetry breaking, the nonlocal interaction of quarks with meson fields generates corresponding quark self-interaction vertices and interaction vertices with gauge fields. These vertices can be incorporated in the model using Schwinger phase factor. The theory of nonlocal interaction  gauge fields with quark and meson fields was constructed in \cite{Terning:1991yt} and later was used to study the nonlocal quark model \cite{Anikin:2000rq,Dorokhov:2015psa, Dorokhov:2011zf,Noguera:2005ej,Villafane:2016ukb}.   

The nonlocal vertex of interaction quark-antiquark with external field can be written as:  
\begin{eqnarray} 
\Gamma_\mu(q)=\gamma_\mu-(p_2+p_1)_\mu m(p_1,p_2),
\end{eqnarray}
where $p_1$ and $p_2=p_1+q$ are momenta of quarks, $q$ is momentum of external field \cite{Anikin:2000rq,Dorokhov:2015psa, Dorokhov:2011zf}. Note that one can generate infinite number interaction vertices of quarks with external gauge fields. For interaction of quark-antiquark with scalar or pseudoscalar mesons, vertices take forms:
\begin{eqnarray} 
&&\Gamma_{\sigma}^a=g_{\sigma}(q^2)\tau^a f(p_1^2)f(p_2^2),\\
&&\Gamma_{\pi}^a=g_{\pi}(q^2)i\gamma_5\tau^a f(p_1^2)f(p_2^2),
\end{eqnarray}
where $p_1$ and $p_2=p_1+q$ are momenta of quarks, $q$ is momentum of a meson, $g_{\sigma}(q^2)$ and $g_{\pi}(q^2)$ are constants which describe a renormalization of scalar or pseudoscalar meson fields accordingly. The constants  $g_{(\sigma,\pi)}(q^2)$ are obtained by solving Bethe-Salpeter equation (BSE) \cite{Noguera:2005ej, Anikin:2000rq, Dorokhov:2011zf} in the ladder approximation. As a result of solving of BSE for a propagator of a meson, the corresponding quark-pion constant can be expressed as
\begin{equation}
\frac{1}{-G+\Pi_{(\sigma,\pi)}(p^2)}=\frac{g^2_{(\sigma,\pi)}(p^2)}{p^2-m^2_{(\sigma,\pi)}},
\end{equation}
where $m_{(\sigma,\pi)}$ is mass of sigma or pion mesons.
On mass-shell pion, quark-pion constant can be fixed \cite{Weinberg:1962hj, Salam:1962ap, Dorokhov:2000gu}
\begin{equation}
\frac{1}{g^2_{(\sigma,\pi)}(m^2_{(\sigma,\pi)})}=\frac{\partial\Pi_{(\sigma,\pi)}(p^2)}{\partial p^2}\Bigg|_{p^2=m^2_{(\sigma,\pi)}},
\end{equation}
where $\Pi_{(\sigma,\pi)}(p^2)$ is polarization operator that is defined above.
Constant $g_{\pi}(q^2)$ has a simple connection with weak pion decay constant through Goldgerger-Treinman (GT) relation in chiral limit of model. 

\begin{figure}[ht]
	\centering
	\includegraphics[trim={0 0cm 0 0cm},clip=true, scale=0.8,keepaspectratio=true]{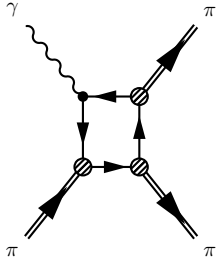}%
	\caption{\label{l1}Feynman diagram which describing transition form-factor  $\gamma^*\pi^-\rightarrow\pi^0\pi^-$. All vertices are nonlocal. }%
\end{figure} 
\section{\label{sec:level3}Form-factor of $F_{3\pi}$}

In this section, the transition form-factor 
photon into three pions in the framework of N$\chi$QM is be described. This form-factor follows from anomalous Ward identity. Here we reproduce the dependence of this form-factor on quark mass in N$\chi$QM model. Note that this dependence is absent in local limit.

In general, the amplitude of  $\gamma^*\rightarrow \pi^+\pi^0\pi^-$ processes can be written as a sum of six diagrams.One of these diagrams is shown on Fig.\ref{l1} and the others can be obtained from it by permutations of pion legs. There are three groups of diagrams which have a symmetry under exchange of direction of quark momentum in quark loop $k\rightarrow -k$.

In framework of nonlocal quark model with linear case of bosonization, contacted nonlocal vertices of interaction pion with quark-quark-photon exist \cite{Dorokhov:2012qa,Dumm:2013zoa,Noguera:2005ej}. These vertices
don't make a contribution in $F_{3\pi}$ transition form factor because terms which contain them vanish after Dirac matrices trace calculation.

 The amplitude of transition of gamma in three pions can be written as
\begin{eqnarray}
A(\gamma\rightarrow \pi^+\pi^0\pi^-)=-iF_{3\pi}(s,t,u)\epsilon^{\mu\nu\alpha\beta}\epsilon^{\mu} p_0^{\nu} p_1^{\alpha}p_2^{\beta},  \quad
\end{eqnarray}
where $p_i$ are momenta of pions, $\epsilon^{\mu}$ is a polarization of photon and $F_{3\pi}(s,t,u)$ is a Lorentz scalar function of the Mandelstam
variables. It can be defined from three types of diagrams in different kinematics:
\begin{eqnarray}
F_{3\pi}(s,t,u)=F_1(s,t,u)+F_2(t,s,u)+F_3(u,t,s),
\end{eqnarray}
where $s,t,u$ - are Mandelstam invariance variables. The first function of transition form-factor can be calculated from the Feynman diagrams Fig.\ref{l1} and in $N\chi QM$ model have a form

\begin{widetext}
\begin{eqnarray}
F_1(s,t,u)=&& e N_c\int\frac{d_E^4 k}{(2\pi)^4}\frac{ g_{\pi}(p_0^2)g_{\pi}(p_1^2)g_{\pi}(p_2^2)f_{k} f_{k+p_1}^2f_{k-p_0}^2f_{k-p_0-p_2}\mathrm{Tr}_f[Q(\tau^-\tau^3\tau^++\tau^+\tau^3\tau^-)]}{D(k)D(k+p_1)D(k-p_0)D(k-p_0-p_2)}\times\nonumber\\
&&\times 4\,\left\{m(k^2)[A+1-B]-m((k-p_0)^2)[C+A] +m((k+p_1)^2)C+m((k-p_0-p_2)^2)B\right\} ,
\label{str_abc}
\end{eqnarray}
	where $D(k)=k^2+m^2(k^2)$, $Q$ is a charge matrix of quark and $\tau^{\pm}=(\tau^1\pm\tau^2)/\sqrt{2}$ are combinations of Pauli matrices that follow from pion fields. A coefficients $A,B,C$ can be found in Appendix A. Structure functions $F_2(s,u,t)$ and $F_3(t,s,u)$ have a connection with $F_1(s,t,u)$ by crossing symmetry, namely:
\begin{equation}
F_2(t,s,u)=F_1(s,t,u) (p_0\leftrightarrows -p_1, \tau^3 \leftrightarrow \tau^+) \qquad \text{and} \qquad F_3(u,t,s)=F_1(s,t,u) (p_0\leftrightarrow -p_2, \tau^3 \leftrightarrow \tau^-).
\end{equation}

	
 	In low energy limit, when kinematic invariants $s=t=u=0$ and performing the chiral expansion over $m_c$ the  value of transition form factor has a form
\begin{eqnarray}
F_{3\pi}(0,0,0)=eN_cN_f g_{\pi}^3\int \frac{d_E^4k}{(2\pi)^4} && f^6(k) \left\{\, 4\left[\frac{(m_0(k^2)
- m_0^{\prime}(k^2)k^2)}{D(k)^4}\right]\right. \nonumber\\
&& \left. -32\,m_c \left[\frac{(m_0^2(k^2)-m_0(k^2)m_0^{\prime}(k^2)k^2)}{D(k)^5}-\frac{1}{8}\frac{1}{D(k)^4}\right]\right\}.
\label{eqFF}
\end{eqnarray}
Here $m(k^2) \to m_0(k^2)=m_df^2(k^2)$, $g_\pi=g_\pi(0)$ and second term in eq.(\ref{eqFF}) has a dependence from current mass of quark. 
In chiral limit current quark mass $m_c$ is zero this form factor takes a form
\begin{eqnarray}
F_{3\pi}(0,0,0)=\frac{eN_cN_f}{f_\pi^3} \int \frac{d_E^4k}{(2\pi)^4}\left[\frac{4m_0^4(k^2)
 - 4m_0^{\prime}(k^2)m_0^3(k^2)k^2}{D(k)^4}\right],
\label{30ll}
\end{eqnarray}
\end{widetext}
where $m_0^{\prime}(k^2)=\dfrac{\partial m_0(k^2)}{\partial k^2}$. Goldgerger-Treinman relation at the quark level which holds in chiral limit of quark model, $f_\pi=g_\pi/m_d$, is used  \cite{Scarpettini:2003fj, Dorokhov:2003kf,Osipov:2007zz}.

When the parameter of nonlocality tends to infinity $\Lambda\rightarrow \infty$, $f(k^2)\rightarrow1$ and $m^{\prime}(k) =0$, $m(k^2)=m_d$  in chiral case. Then the integral in Eq.(\ref{30ll}) can be solved analytically:
\begin{eqnarray}
\int_{0}^{\infty}dk^2 \frac{k^2m_d^4}{(k^2+m_d^2)^4}=\frac{1}{6}.
\end{eqnarray}
In this particular limit, the eq.(\ref{30ll}) reproduces  WZW form-factor \cite{Wess:1971yu}: 
\begin{eqnarray}
F_{3\pi}=\dfrac{eN_cN_f}{24\pi^2f_\pi^3}=\dfrac{e}{4\pi^2f_\pi^3}\simeq9.72 \,(0.09) \, \mathrm{GeV}^{-3}.
\end{eqnarray}

\begin{figure}[h!]
	\includegraphics[ scale=0.34]{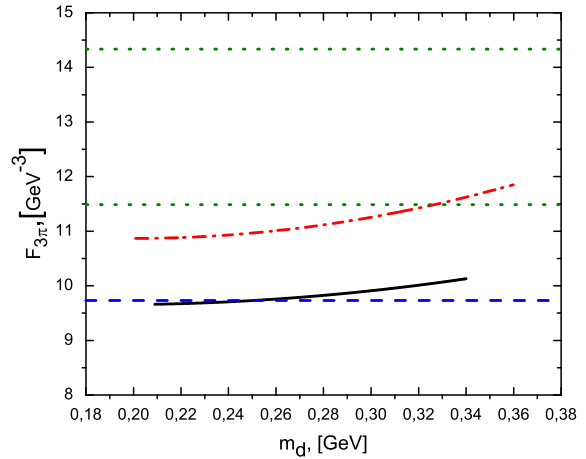}%
	\caption{\label{l2} Dependence of transition form-factor  $\gamma^*\pi^-\rightarrow\pi^0\pi^-$ on dynamical mass of quark $m_d$. Solid black line show a behavior of $F_{3\pi}$ in chiral limit, dash-dot red line corresponds to the transition form-factor on the physical threshold of process. The area between the olive dot lines show a range from Serpukhov experiment \cite{Antipov:1986tp}. Blue dashed line corresponds a limit value $F_{3\pi}$ when quark-meson nonlocality form factor tends to zero, $f(p^2) \to 0$, in N$\chi$QM model or quantity obtained from low-energy theorem.}
\end{figure}

 Behavior of the form factor $F_{3\pi}$ in N$\chi$QM model is different on pion mass-shell and in chiral limit due to nonzero current quark mass. 
The dependence of $F_{3\pi}$ form-factoron dynamical quark mass for both cases of pion mass-shell and chiral limit is shown on  Fig.\ref{l2}. Averaged quantity of $F_{3\pi}$ at low energy in chiral limit is
\begin{eqnarray}
	F_{3\pi}(0,0,0)=9.789 \,(0.4) \, \mathrm{GeV}^{-3}.
\end{eqnarray}

Therefore, we can conclude that the presence of nonlocality does not change the picture in chiral limit and the obtained value of $F_{3\pi}$ is in agreement with the one obtained from the low-energy theorem.

For physical masses of pions, transition form factor should be calculated on the physical threshold for $q^2=0$ and $s+t+u=3m_\pi^2$. In this case,
kinematics variables \cite{Giller:2005uy} take the form $s^{thr}=(m_{\pi-}+m_{\pi^0})^2$, $t^{thr}=-m_{\pi^-}m_{\pi^0}^2/(m_{\pi^-}+m_{\pi^0})$ and $u^{thr}=m_{\pi^-}(m_{\pi^-}^2-m_{\pi^-}m_{\pi^0}-m_{\pi^0}^2)/(m_{\pi^-}+m_{\pi^0})$. Thus in lowest order of perturbation by $m_\pi^2$ transition form factor $F_{3\pi}^{thr}$ has the form:
\begin{eqnarray}
F_{3\pi}^{thr}&&(s^{thr},t^{thr},u^{thr})=eN_cN_fg_\pi^3(m_\pi^2)\int \frac{d_E^4k}{(2\pi)^4}f^6(k^2) \times\nonumber\\
&&	\times\left[\frac{4m(k^2)-4m^{\prime}(k^2)k^2}{D(k)^4}\right]+\mathcal{O}(m_\pi^2).
\label{phys_ll}
\end{eqnarray}
Note that this form is similar to the one obtained in chiral limit. The corrections due to pion mass are suppressed. Dependence on current quark mass leads to a following change of transition form factor:
\begin{eqnarray}
F_{3\pi}^{thr}(s^{thr},t^{thr},u^{thr})=11.48 \,(0.58) \, \mathrm{GeV}^{-3}.
\end{eqnarray}
Direct violation of chiral symmetry by including nonzero current quark mass leads to increase of value of transition form factor $F_{3\pi}$  (see Fig.\ref{l2}). The shift of decay constant of pion at nonzero current mass of quark plays an important role here. 
 
As in the case of pion transition form factor into two photons \cite{Bernstein:2011bx}, $F_{3\pi}$ form factor is slowly changed with the growth of dynamical mass of quark. Inclusion of current mass of quark is a more important effect for the model. 
 In lowest order of decomposition in powers of electromagnetic constant, nonlocal interaction external electromagnetic fields with quarks does not play a significant roleand the important contribution is due to local vertex of interaction of quarks with a photon. Same effect occurs for transition form factor of pion into two photons \cite{Dorokhov:2011zf}. Nonlocality of meson--quarks vertices provided by existence of form factors $f(p^2)$. The model in chiral limit reproduce the form of expression  $F_{3\pi}$ that follows from the WZW action. The  value $F_{3\pi}$ in N$\chi$QM in chiral limit is in agreement with the one from low energy theorem.

 	
 	\section{\label{sec:Outlook}Outlook}
 	
	 	In article the transition form-factor of  $\gamma^*\rightarrow\pi^+\pi^0\pi^-$ which reproduce a WZW term from anomalous Ward identities in chiral limit of model was studied. It is shown that the nonlocality does not play a significant role for the form factor $F_{3\pi}$ in chiral limit. on the physical threshold value transition form factor in nonlocal quark model $F_{3\pi}^{thr}=11.48 \,(0.58) \, \mathrm{GeV}^{-3}$ has a good agreement with experimental which was measured by Antipov et al. \cite{Antipov:1986tp}. The dynamical quark mass has in the typical interval $200 \div 350$ MeV for physical mass, and the current mass of quark has a typical values in range of $4 \div 9$ MeV.
	 	
	 	  The result obtained for $F_{3\pi}$ has an agreement with results from dispersion analysis in two-loop \cite{Hannah:2001ee}, in framework of integral equation approach \cite{Truong:2001en} and from the case of dispersive representations with the $\pi\pi$ P-wave phase 
	 	 shift \cite{Hoferichter:2012pm}. ChPT also has an agreement with experimental data and with this result in case when take into account correction at $O(p^6)$ with $q^2$ - dependence and electromagnetic corrections	\cite{Ametller:2001yk}. Full agreement with this result have ChPT in case of $O(p^8)$ when electromagnetic corrections at $O(e^2)$ are included. The calculations in N$\chi$QM coincide with ladder Dyson-Schwinger calculation \cite{Cotanch:2003xv} which also take into account the dependence of mass on momentum of quark. One can say that nonperturbation effects are incorporated inside N$\chi$QM by inclusion of nonlocality structure of interaction quarks with mesons.   

	Recently it was shown that picture of calculation will change if one takes into account of intermediate vector meson resonance exchange \cite{Briceno:2015dca,Briceno:2016kkp}. Similar picture should appear in the processes $\pi\gamma^* \rightarrow \pi\pi/K\bar{K}$. We plan to consider a role lightest vector meson in these reactions in framework of extended $SU(2) \times SU(2)$ model \cite{Dorokhov:2003kf, Villafane:2016ukb}. It is also interesting in connection with future measurement of exclusive $\gamma p \rightarrow \pi^+\pi^- p$ reaction of photoproduction at the GlueX experiment at Jefferson laboratory \cite{Ghoul:2015ifw}.
		
In future we plan to calculate Dalitz decays of $\eta$ and $\eta^\prime$ mesons into $\gamma\pi^+\pi^-$ or $e^+e^-\pi^+\pi^-$. These decays are well-studied experimentally and will be measured in future experiments with very high accuracy. The important role in these processes plays vector mesons which have to be included in extended $SU(3) \times SU(3)$ model. 

\begin{acknowledgments}
	 I thank A.~E.~Dorokhov, A.~E.~Radzhabov, D.~V.~Karlovets and A.~A.~Shishmarev for discussion and for close reading. I thankful to M.~V.~Polyakov for motivation to study this subject.
The work is supported by Russian Science Foundation grant (RSCF 15-12-10009). 	
\end{acknowledgments}	

	
	%
	%
	
	%
	
	

\appendix

\section{Kinematic functions}

Functions A, B, C  that appeared in eq.(\ref{str_abc}) after rewriting dependence of Lorenz-index loop integral in terms of external fields have the forms: 
\begin{eqnarray}
 &&A=\frac{(kp_1)(p_0^2p_2^2-(p_0p_2)^2)}{RT}\nonumber\\&&+\frac{(kp_2)((p_0p_1)(p_0p_2)-p_0^2(p_1p_2))}{RT}\nonumber\\
 &&+\frac{(kp_0)((p_0p_2)(p_1p_2)-(p_0p_1)p_2^2))}{RT},
\end{eqnarray}
\begin{eqnarray}
&&B=\frac{(kp_1)((p_0p_2)(p_0p_1)-p_0^2(p_1p_2))}{RT}\nonumber\\&&+\frac{(kp_2)(p_0^2p_1^2-(p_0p_1)^2)}{RT}\nonumber\\
&&+\frac{(kp_0)((p_0p_1)(p_1p_2)-(p_0p_2)p_1^2))}{RT},
\end{eqnarray}
\begin{eqnarray}
&&C=\frac{(kp_1)((p_0p_2)(p_1p_2)-(p_0p_1)p_2^2)}{RT}\nonumber\\&&+\frac{(kp_2)((p_0p_1)(p_1p_2)-p_1^2(p_0p_2))}{RT}\nonumber\\
&&+\frac{(kp_0)(p_1^2p_2^2-(p_1p_2)^2))}{RT},
\end{eqnarray}

where denominator $RT$ is
\begin{eqnarray}
RT=&&p_0^2p_1^2p_2^2+2(p_0p_1)(p_0p_2)(p_1p_2)\nonumber\\
&&-p_0^2(p_1p_2)^2-(p_0p_1)^2p_2^2-(p_0p_2)^2p_1^2.
\end{eqnarray}

These coefficients have a crossing symmetry: $C=A(p_0\leftrightarrow -p_1)$ and $B=A(p_0\leftrightarrow -p_2)$.


\begin{thebibliography}{47}%
	\makeatletter
	\providecommand \@ifxundefined [1]{%
		\@ifx{#1\undefined}
	}%
	\providecommand \@ifnum [1]{%
		\ifnum #1\expandafter \@firstoftwo
		\else \expandafter \@secondoftwo
		\fi
	}%
	\providecommand \@ifx [1]{%
		\ifx #1\expandafter \@firstoftwo
		\else \expandafter \@secondoftwo
		\fi
	}%
	\providecommand \natexlab [1]{#1}%
	\providecommand \enquote  [1]{``#1''}%
	\providecommand \bibnamefont  [1]{#1}%
	\providecommand \bibfnamefont [1]{#1}%
	\providecommand \citenamefont [1]{#1}%
	\providecommand \href@noop [0]{\@secondoftwo}%
	\providecommand \href [0]{\begingroup \@sanitize@url \@href}%
	\providecommand \@href[1]{\@@startlink{#1}\@@href}%
	\providecommand \@@href[1]{\endgroup#1\@@endlink}%
	\providecommand \@sanitize@url [0]{\catcode `\\12\catcode `\$12\catcode
		`\&12\catcode `\#12\catcode `\^12\catcode `\_12\catcode `\%12\relax}%
	\providecommand \@@startlink[1]{}%
	\providecommand \@@endlink[0]{}%
	\providecommand \url  [0]{\begingroup\@sanitize@url \@url }%
	\providecommand \@url [1]{\endgroup\@href {#1}{\urlprefix }}%
	\providecommand \urlprefix  [0]{URL }%
	\providecommand \Eprint [0]{\href }%
	\providecommand \doibase [0]{http://dx.doi.org/}%
	\providecommand \selectlanguage [0]{\@gobble}%
	\providecommand \bibinfo  [0]{\@secondoftwo}%
	\providecommand \bibfield  [0]{\@secondoftwo}%
	\providecommand \translation [1]{[#1]}%
	\providecommand \BibitemOpen [0]{}%
	\providecommand \bibitemStop [0]{}%
	\providecommand \bibitemNoStop [0]{.\EOS\space}%
	\providecommand \EOS [0]{\spacefactor3000\relax}%
	\providecommand \BibitemShut  [1]{\csname bibitem#1\endcsname}%
	\let\auto@bib@innerbib\@empty
	\bibitem [{\citenamefont {Steinberger}(1949)}]{Steinberger:1949wx}%
	\BibitemOpen
	\bibfield  {author} {\bibinfo {author} {\bibfnamefont {J.}~\bibnamefont
			{Steinberger}},\ }\href@noop {} {\bibfield  {journal} {\bibinfo  {journal}
			{Phys.\ Rev.}\ }\textbf {\bibinfo {volume} {76}},\ \bibinfo {pages} {1180}
		(\bibinfo {year} {1949})}\BibitemShut {NoStop}%
	\bibitem [{\citenamefont {Wess}\ and\ \citenamefont
		{Zumino}(1971)}]{Wess:1971yu}%
	\BibitemOpen
	\bibfield  {author} {\bibinfo {author} {\bibfnamefont {J.}~\bibnamefont
			{Wess}}\ and\ \bibinfo {author} {\bibfnamefont {B.}~\bibnamefont {Zumino}},\
	}\href@noop {} {\bibfield  {journal} {\bibinfo  {journal} {Phys.\ Lett.}\
	}\textbf {\bibinfo {volume} {37B}},\ \bibinfo {pages} {95} (\bibinfo {year}
	{1971})}\BibitemShut {NoStop}%
\bibitem [{\citenamefont {Witten}(1983)}]{Witten:1983tw}%
\BibitemOpen
\bibfield  {author} {\bibinfo {author} {\bibfnamefont {E.}~\bibnamefont
		{Witten}},\ }\href@noop {} {\bibfield  {journal} {\bibinfo  {journal} {Nucl.\
			Phys.\ B}\ }\textbf {\bibinfo {volume} {223}},\ \bibinfo {pages} {422}
	(\bibinfo {year} {1983})}\BibitemShut {NoStop}%
\bibitem [{\citenamefont {Ruiz~Arriola}\ and\ \citenamefont
	{Broniowski}(2010)}]{Arriola:2010aq}%
\BibitemOpen
\bibfield  {author} {\bibinfo {author} {\bibfnamefont {E.}~\bibnamefont
		{Ruiz~Arriola}}\ and\ \bibinfo {author} {\bibfnamefont {W.}~\bibnamefont
		{Broniowski}},\ }\href@noop {} {\bibfield  {journal} {\bibinfo  {journal}
		{Phys. Rev.}\ }\textbf {\bibinfo {volume} {D81}},\ \bibinfo {pages} {094021}
	(\bibinfo {year} {2010})}\BibitemShut {NoStop}%
\bibitem [{\citenamefont {Knecht}\ and\ \citenamefont
	{Nyffeler}(2002)}]{Knecht:2001qf}%
\BibitemOpen
\bibfield  {author} {\bibinfo {author} {\bibfnamefont {M.}~\bibnamefont
		{Knecht}}\ and\ \bibinfo {author} {\bibfnamefont {A.}~\bibnamefont
		{Nyffeler}},\ }\href@noop {} {\bibfield  {journal} {\bibinfo  {journal}
		{Phys. Rev.}\ }\textbf {\bibinfo {volume} {D65}},\ \bibinfo {pages} {073034}
	(\bibinfo {year} {2002})}\BibitemShut {NoStop}%
\bibitem [{\citenamefont {Mikhailov}\ \emph {et~al.}(2010)\citenamefont
	{Mikhailov}, \citenamefont {Pimikov},\ and\ \citenamefont
	{Stefanis}}]{Mikhailov:2010ud}%
\BibitemOpen
\bibfield  {author} {\bibinfo {author} {\bibfnamefont {S.~V.}\ \bibnamefont
		{Mikhailov}}, \bibinfo {author} {\bibfnamefont {A.~V.}\ \bibnamefont
		{Pimikov}}, \ and\ \bibinfo {author} {\bibfnamefont {N.~G.}\ \bibnamefont
		{Stefanis}},\ }\href@noop {} {\bibfield  {journal} {\bibinfo  {journal}
		{Phys. Rev.}\ }\textbf {\bibinfo {volume} {D82}},\ \bibinfo {pages} {054020}
	(\bibinfo {year} {2010})}\BibitemShut {NoStop}%
\bibitem [{\citenamefont {Oganesian}\ \emph {et~al.}(2016)\citenamefont
	{Oganesian}, \citenamefont {Pimikov}, \citenamefont {Stefanis},\ and\
	\citenamefont {Teryaev}}]{Oganesian:2015ucv}%
\BibitemOpen
\bibfield  {author} {\bibinfo {author} {\bibfnamefont {A.~G.}\ \bibnamefont
		{Oganesian}}, \bibinfo {author} {\bibfnamefont {A.~V.}\ \bibnamefont
		{Pimikov}}, \bibinfo {author} {\bibfnamefont {N.~G.}\ \bibnamefont
		{Stefanis}}, \ and\ \bibinfo {author} {\bibfnamefont {O.~V.}\ \bibnamefont
		{Teryaev}},\ }\href@noop {} {\bibfield  {journal} {\bibinfo  {journal} {Phys.
			Rev.}\ }\textbf {\bibinfo {volume} {D93}},\ \bibinfo {pages} {054040}
	(\bibinfo {year} {2016})}\BibitemShut {NoStop}%
\bibitem [{\citenamefont {Dorokhov}\ \emph {et~al.}(2011)\citenamefont
	{Dorokhov}, \citenamefont {Radzhabov},\ and\ \citenamefont
	{Zhevlakov}}]{Dorokhov:2011zf}%
\BibitemOpen
\bibfield  {author} {\bibinfo {author} {\bibfnamefont {A.~E.}\ \bibnamefont
		{Dorokhov}}, \bibinfo {author} {\bibfnamefont {A.~E.}\ \bibnamefont
		{Radzhabov}}, \ and\ \bibinfo {author} {\bibfnamefont {A.}~\bibnamefont
		{Zhevlakov}},\ }\href@noop {} {\bibfield  {journal} {\bibinfo  {journal}
		{Eur. Phys. J.}\ }\textbf {\bibinfo {volume} {C71}},\ \bibinfo {pages} {1702}
	(\bibinfo {year} {2011})}\BibitemShut {NoStop}%
\bibitem [{\citenamefont {Dumm}\ \emph {et~al.}(2014)\citenamefont {Dumm},
	\citenamefont {Noguera}, \citenamefont {Scoccola},\ and\ \citenamefont
	{Scopetta}}]{Dumm:2013zoa}%
\BibitemOpen
\bibfield  {author} {\bibinfo {author} {\bibfnamefont {D.~G.}\ \bibnamefont
		{Dumm}}, \bibinfo {author} {\bibfnamefont {S.}~\bibnamefont {Noguera}},
	\bibinfo {author} {\bibfnamefont {N.~N.}\ \bibnamefont {Scoccola}}, \ and\
	\bibinfo {author} {\bibfnamefont {S.}~\bibnamefont {Scopetta}},\ }\href
{\doibase 10.1103/PhysRevD.89.054031} {\bibfield  {journal} {\bibinfo
		{journal} {Phys. Rev.}\ }\textbf {\bibinfo {volume} {D89}},\ \bibinfo {pages}
	{054031} (\bibinfo {year} {2014})}\BibitemShut {NoStop}%
\bibitem [{\citenamefont {Bernstein}\ and\ \citenamefont
	{Holstein}(2013)}]{Bernstein:2011bx}%
\BibitemOpen
\bibfield  {author} {\bibinfo {author} {\bibfnamefont {A.~M.}\ \bibnamefont
		{Bernstein}}\ and\ \bibinfo {author} {\bibfnamefont {B.~R.}\ \bibnamefont
		{Holstein}},\ }\href@noop {} {\bibfield  {journal} {\bibinfo  {journal}
		{Rev.\ Mod.\ Phys.}\ }\textbf {\bibinfo {volume} {85}},\ \bibinfo {pages}
	{49} (\bibinfo {year} {2013})}\BibitemShut {NoStop}%
\bibitem [{\citenamefont {Bijnens}(1991)}]{Bijnens:1991db}%
\BibitemOpen
\bibfield  {author} {\bibinfo {author} {\bibfnamefont {J.}~\bibnamefont
		{Bijnens}},\ }\href@noop {} {\bibfield  {journal} {\bibinfo  {journal}
		{Nucl.\ Phys.\ B}\ }\textbf {\bibinfo {volume} {367}},\ \bibinfo {pages}
	{709} (\bibinfo {year} {1991})}\BibitemShut {NoStop}%
\bibitem [{\citenamefont {Giller}\ \emph {et~al.}(2005)\citenamefont {Giller},
	\citenamefont {Ocherashvili}, \citenamefont {Ebertshauser}, \citenamefont
	{Moinester},\ and\ \citenamefont {Scherer}}]{Giller:2005uy}%
\BibitemOpen
\bibfield  {author} {\bibinfo {author} {\bibfnamefont {I.}~\bibnamefont
		{Giller}}, \bibinfo {author} {\bibfnamefont {A.}~\bibnamefont
		{Ocherashvili}}, \bibinfo {author} {\bibfnamefont {T.}~\bibnamefont
		{Ebertshauser}}, \bibinfo {author} {\bibfnamefont {M.~A.}\ \bibnamefont
		{Moinester}}, \ and\ \bibinfo {author} {\bibfnamefont {S.}~\bibnamefont
		{Scherer}},\ }\href@noop {} {\bibfield  {journal} {\bibinfo  {journal} {Eur.
			Phys. J.}\ }\textbf {\bibinfo {volume} {A25}},\ \bibinfo {pages} {229}
	(\bibinfo {year} {2005})}\BibitemShut {NoStop}%
\bibitem [{\citenamefont {Antipov}\ \emph {et~al.}(1987)\citenamefont {Antipov}
	\emph {et~al.}}]{Antipov:1986tp}%
\BibitemOpen
\bibfield  {author} {\bibinfo {author} {\bibfnamefont {{\relax Yu}.~M.}\
		\bibnamefont {Antipov}} \emph {et~al.},\ }\href@noop {} {\bibfield  {journal}
	{\bibinfo  {journal} {Phys. Rev.}\ }\textbf {\bibinfo {volume} {D36}},\
	\bibinfo {pages} {21} (\bibinfo {year} {1987})}\BibitemShut {NoStop}%
\bibitem [{\citenamefont {et~al}(1996)}]{CERN_Baum}%
\BibitemOpen
\bibfield  {author} {\bibinfo {author} {\bibfnamefont {G.~Baum.}\ \bibnamefont
		{et~al}},\ }\href@noop {} {\bibfield  {journal} {\bibinfo  {journal}
		{CERN-SPSLC}\ }\textbf {\bibinfo {volume} {96-14}} (\bibinfo {year}
	{1996})}\BibitemShut {NoStop}%
\bibitem [{\citenamefont {Ametller}\ \emph {et~al.}(2001)\citenamefont
	{Ametller}, \citenamefont {Knecht},\ and\ \citenamefont
	{Talavera}}]{Ametller:2001yk}%
\BibitemOpen
\bibfield  {author} {\bibinfo {author} {\bibfnamefont {L.}~\bibnamefont
		{Ametller}}, \bibinfo {author} {\bibfnamefont {M.}~\bibnamefont {Knecht}}, \
	and\ \bibinfo {author} {\bibfnamefont {P.}~\bibnamefont {Talavera}},\
}\href@noop {} {\bibfield  {journal} {\bibinfo  {journal} {Phys. Rev.}\
}\textbf {\bibinfo {volume} {D64}},\ \bibinfo {pages} {094009} (\bibinfo
{year} {2001})}\BibitemShut {NoStop}%
\bibitem [{\citenamefont {Anikin}\ \emph {et~al.}(2000)\citenamefont {Anikin},
	\citenamefont {Dorokhov},\ and\ \citenamefont {Tomio}}]{Anikin:2000rq}%
\BibitemOpen
\bibfield  {author} {\bibinfo {author} {\bibfnamefont {I.~V.}\ \bibnamefont
		{Anikin}}, \bibinfo {author} {\bibfnamefont {A.~E.}\ \bibnamefont
		{Dorokhov}}, \ and\ \bibinfo {author} {\bibfnamefont {L.}~\bibnamefont
		{Tomio}},\ }\href@noop {} {\bibfield  {journal} {\bibinfo  {journal} {Phys.
			Part. Nucl.}\ }\textbf {\bibinfo {volume} {31}},\ \bibinfo {pages} {509}
	(\bibinfo {year} {2000})},\ \bibinfo {note} {[Fiz. Elem. Chast. Atom.
	Yadra31,1023(2000)]}\BibitemShut {NoStop}%
\bibitem [{\citenamefont {Dorokhov}\ and\ \citenamefont
	{Tomio}(2000)}]{Dorokhov:2000gu}%
\BibitemOpen
\bibfield  {author} {\bibinfo {author} {\bibfnamefont {A.~E.}\ \bibnamefont
		{Dorokhov}}\ and\ \bibinfo {author} {\bibfnamefont {L.}~\bibnamefont
		{Tomio}},\ }\href {\doibase 10.1103/PhysRevD.62.014016} {\bibfield  {journal}
	{\bibinfo  {journal} {Phys. Rev.}\ }\textbf {\bibinfo {volume} {D62}},\
	\bibinfo {pages} {014016} (\bibinfo {year} {2000})}\BibitemShut {NoStop}%
\bibitem [{\citenamefont {Noguera}(2007)}]{Noguera:2005ej}%
\BibitemOpen
\bibfield  {author} {\bibinfo {author} {\bibfnamefont {S.}~\bibnamefont
		{Noguera}},\ }\href {\doibase 10.1142/S021830130700565X} {\bibfield
	{journal} {\bibinfo  {journal} {Int. J. Mod. Phys.}\ }\textbf {\bibinfo
		{volume} {E16}},\ \bibinfo {pages} {97} (\bibinfo {year} {2007})}\BibitemShut
{NoStop}%
\bibitem [{\citenamefont {Noguera}\ and\ \citenamefont
	{Scoccola}(2008)}]{Noguera:2008cm}%
\BibitemOpen
\bibfield  {author} {\bibinfo {author} {\bibfnamefont {S.}~\bibnamefont
		{Noguera}}\ and\ \bibinfo {author} {\bibfnamefont {N.~N.}\ \bibnamefont
		{Scoccola}},\ }\href {\doibase 10.1103/PhysRevD.78.114002} {\bibfield
	{journal} {\bibinfo  {journal} {Phys. Rev.}\ }\textbf {\bibinfo {volume}
		{D78}},\ \bibinfo {pages} {114002} (\bibinfo {year} {2008})}\BibitemShut
{NoStop}%
\bibitem [{\citenamefont {Scarpettini}\ \emph {et~al.}(2004)\citenamefont
	{Scarpettini}, \citenamefont {Gomez~Dumm},\ and\ \citenamefont
	{Scoccola}}]{Scarpettini:2003fj}%
\BibitemOpen
\bibfield  {author} {\bibinfo {author} {\bibfnamefont {A.}~\bibnamefont
		{Scarpettini}}, \bibinfo {author} {\bibfnamefont {D.}~\bibnamefont
		{Gomez~Dumm}}, \ and\ \bibinfo {author} {\bibfnamefont {N.~N.}\ \bibnamefont
		{Scoccola}},\ }\href@noop {} {\bibfield  {journal} {\bibinfo  {journal}
		{Phys. Rev.}\ }\textbf {\bibinfo {volume} {D69}},\ \bibinfo {pages} {114018}
	(\bibinfo {year} {2004})}\BibitemShut {NoStop}%
\bibitem [{\citenamefont {Radzhabov}\ and\ \citenamefont
	{Volkov}(2004)}]{Radzhabov:2003hy}%
\BibitemOpen
\bibfield  {author} {\bibinfo {author} {\bibfnamefont {A.~E.}\ \bibnamefont
		{Radzhabov}}\ and\ \bibinfo {author} {\bibfnamefont {M.~K.}\ \bibnamefont
		{Volkov}},\ }\href {\doibase 10.1140/epja/i2003-10112-5} {\bibfield
	{journal} {\bibinfo  {journal} {Eur. Phys. J.}\ }\textbf {\bibinfo {volume}
		{A19}},\ \bibinfo {pages} {139} (\bibinfo {year} {2004})},\ \bibinfo {note}
{[Phys. Part. Nucl. Lett.1,1(2004)]}\BibitemShut {NoStop}%
\bibitem [{\citenamefont {Radzhabov}\ \emph {et~al.}(2011)\citenamefont
	{Radzhabov}, \citenamefont {Blaschke}, \citenamefont {Buballa},\ and\
	\citenamefont {Volkov}}]{Radzhabov:2010dd}%
\BibitemOpen
\bibfield  {author} {\bibinfo {author} {\bibfnamefont {A.~E.}\ \bibnamefont
		{Radzhabov}}, \bibinfo {author} {\bibfnamefont {D.}~\bibnamefont {Blaschke}},
	\bibinfo {author} {\bibfnamefont {M.}~\bibnamefont {Buballa}}, \ and\
	\bibinfo {author} {\bibfnamefont {M.~K.}\ \bibnamefont {Volkov}},\ }\href
{\doibase 10.1103/PhysRevD.83.116004} {\bibfield  {journal} {\bibinfo
		{journal} {Phys. Rev.}\ }\textbf {\bibinfo {volume} {D83}},\ \bibinfo {pages}
	{116004} (\bibinfo {year} {2011})}\BibitemShut {NoStop}%
\bibitem [{\citenamefont {Gomez~Dumm}\ \emph
	{et~al.}(2006{\natexlab{a}})\citenamefont {Gomez~Dumm}, \citenamefont
	{Blaschke}, \citenamefont {Grunfeld},\ and\ \citenamefont
	{Scoccola}}]{GomezDumm:2005hy}%
\BibitemOpen
\bibfield  {author} {\bibinfo {author} {\bibfnamefont {D.}~\bibnamefont
		{Gomez~Dumm}}, \bibinfo {author} {\bibfnamefont {D.~B.}\ \bibnamefont
		{Blaschke}}, \bibinfo {author} {\bibfnamefont {A.~G.}\ \bibnamefont
		{Grunfeld}}, \ and\ \bibinfo {author} {\bibfnamefont {N.~N.}\ \bibnamefont
		{Scoccola}},\ }\href {\doibase 10.1103/PhysRevD.73.114019} {\bibfield
	{journal} {\bibinfo  {journal} {Phys. Rev.}\ }\textbf {\bibinfo {volume}
		{D73}},\ \bibinfo {pages} {114019} (\bibinfo {year}
	{2006}{\natexlab{a}})}\BibitemShut {NoStop}%
\bibitem [{Note1()}]{Note1}%
\BibitemOpen
\bibinfo {note} {We consider the isospin limit where masses of quarks $u$ and
	$d$ are equal.}\BibitemShut {Stop}%
\bibitem [{\citenamefont {Izzo~Villafañe}\ \emph {et~al.}(2016)\citenamefont
	{Izzo~Villafañe}, \citenamefont {Gomez~Dumm},\ and\ \citenamefont
	{Scoccola}}]{Villafane:2016ukb}%
\BibitemOpen
\bibfield  {author} {\bibinfo {author} {\bibfnamefont {M.}~\bibnamefont
		{Izzo~Villafañe}}, \bibinfo {author} {\bibfnamefont {D.}~\bibnamefont
		{Gomez~Dumm}}, \ and\ \bibinfo {author} {\bibfnamefont {N.~N.}\ \bibnamefont
		{Scoccola}},\ }\href {\doibase 10.1103/PhysRevD.94.054003} {\bibfield
	{journal} {\bibinfo  {journal} {Phys. Rev.}\ }\textbf {\bibinfo {volume}
		{D94}},\ \bibinfo {pages} {054003} (\bibinfo {year} {2016})}\BibitemShut
{NoStop}%
\bibitem [{\citenamefont {Anikin}\ \emph {et~al.}(1994)\citenamefont {Anikin},
	\citenamefont {Ivanov}, \citenamefont {Kulimanova},\ and\ \citenamefont
	{Lyubovitskij}}]{Anikin:1993mm}%
\BibitemOpen
\bibfield  {author} {\bibinfo {author} {\bibfnamefont {I.}~\bibnamefont
		{Anikin}}, \bibinfo {author} {\bibfnamefont {M.~A.}\ \bibnamefont {Ivanov}},
	\bibinfo {author} {\bibfnamefont {N.}~\bibnamefont {Kulimanova}}, \ and\
	\bibinfo {author} {\bibfnamefont {V.~E.}\ \bibnamefont {Lyubovitskij}},\
}\href@noop {} {\bibfield  {journal} {\bibinfo  {journal} {Phys. Atom.
		Nucl.}\ }\textbf {\bibinfo {volume} {57}},\ \bibinfo {pages} {1021} (\bibinfo
{year} {1994})},\ \bibinfo {note} {[Yad. Fiz.57,1082(1994)]}\BibitemShut
{NoStop}%
\bibitem [{\citenamefont {Gomez~Dumm}\ \emph
	{et~al.}(2006{\natexlab{b}})\citenamefont {Gomez~Dumm}, \citenamefont
	{Grunfeld},\ and\ \citenamefont {Scoccola}}]{GomezDumm:2006vz}%
\BibitemOpen
\bibfield  {author} {\bibinfo {author} {\bibfnamefont {D.}~\bibnamefont
		{Gomez~Dumm}}, \bibinfo {author} {\bibfnamefont {A.~G.}\ \bibnamefont
		{Grunfeld}}, \ and\ \bibinfo {author} {\bibfnamefont {N.~N.}\ \bibnamefont
		{Scoccola}},\ }\href {\doibase 10.1103/PhysRevD.74.054026} {\bibfield
	{journal} {\bibinfo  {journal} {Phys. Rev.}\ }\textbf {\bibinfo {volume}
		{D74}},\ \bibinfo {pages} {054026} (\bibinfo {year}
	{2006}{\natexlab{b}})}\BibitemShut {NoStop}%
\bibitem [{\citenamefont {Shifman}\ \emph {et~al.}(1979)\citenamefont
	{Shifman}, \citenamefont {Vainshtein},\ and\ \citenamefont
	{Zakharov}}]{Shifman:1978bx}%
\BibitemOpen
\bibfield  {author} {\bibinfo {author} {\bibfnamefont {M.~A.}\ \bibnamefont
		{Shifman}}, \bibinfo {author} {\bibfnamefont {A.~I.}\ \bibnamefont
		{Vainshtein}}, \ and\ \bibinfo {author} {\bibfnamefont {V.~I.}\ \bibnamefont
		{Zakharov}},\ }\href@noop {} {\bibfield  {journal} {\bibinfo  {journal}
		{Nucl. Phys.}\ }\textbf {\bibinfo {volume} {B147}},\ \bibinfo {pages} {385}
	(\bibinfo {year} {1979})}\BibitemShut {NoStop}%
\bibitem [{\citenamefont {Dosch}\ and\ \citenamefont
	{Narison}(1998)}]{Dosch:1997wb}%
\BibitemOpen
\bibfield  {author} {\bibinfo {author} {\bibfnamefont {H.~G.}\ \bibnamefont
		{Dosch}}\ and\ \bibinfo {author} {\bibfnamefont {S.}~\bibnamefont
		{Narison}},\ }\href {\doibase 10.1016/S0370-2693(97)01370-1} {\bibfield
	{journal} {\bibinfo  {journal} {Phys. Lett.}\ }\textbf {\bibinfo {volume}
		{B417}},\ \bibinfo {pages} {173} (\bibinfo {year} {1998})}\BibitemShut
{NoStop}%
\bibitem [{\citenamefont {Giusti}\ \emph {et~al.}(1999)\citenamefont {Giusti},
	\citenamefont {Rapuano}, \citenamefont {Talevi},\ and\ \citenamefont
	{Vladikas}}]{Giusti:1998wy}%
\BibitemOpen
\bibfield  {author} {\bibinfo {author} {\bibfnamefont {L.}~\bibnamefont
		{Giusti}}, \bibinfo {author} {\bibfnamefont {F.}~\bibnamefont {Rapuano}},
	\bibinfo {author} {\bibfnamefont {M.}~\bibnamefont {Talevi}}, \ and\ \bibinfo
	{author} {\bibfnamefont {A.}~\bibnamefont {Vladikas}},\ }\href {\doibase
	10.1016/S0550-3213(98)00659-2} {\bibfield  {journal} {\bibinfo  {journal}
		{Nucl. Phys.}\ }\textbf {\bibinfo {volume} {B538}},\ \bibinfo {pages} {249}
	(\bibinfo {year} {1999})}\BibitemShut {NoStop}%
\bibitem [{\citenamefont {Aoki}\ \emph {et~al.}(2014)\citenamefont {Aoki} \emph
	{et~al.}}]{Aoki:2013ldr}%
\BibitemOpen
\bibfield  {author} {\bibinfo {author} {\bibfnamefont {S.}~\bibnamefont
		{Aoki}} \emph {et~al.},\ }\href {\doibase 10.1140/epjc/s10052-014-2890-7}
{\bibfield  {journal} {\bibinfo  {journal} {Eur. Phys. J.}\ }\textbf
	{\bibinfo {volume} {C74}},\ \bibinfo {pages} {2890} (\bibinfo {year}
	{2014})}\BibitemShut {NoStop}%
\bibitem [{\citenamefont {Colangelo}\ \emph {et~al.}(2011)\citenamefont
	{Colangelo} \emph {et~al.}}]{Colangelo:2010et}%
\BibitemOpen
\bibfield  {author} {\bibinfo {author} {\bibfnamefont {G.}~\bibnamefont
		{Colangelo}} \emph {et~al.},\ }\href {\doibase
	10.1140/epjc/s10052-011-1695-1} {\bibfield  {journal} {\bibinfo  {journal}
		{Eur. Phys. J.}\ }\textbf {\bibinfo {volume} {C71}},\ \bibinfo {pages} {1695}
	(\bibinfo {year} {2011})}\BibitemShut {NoStop}%
\bibitem [{\citenamefont {Kneur}\ and\ \citenamefont
	{Neveu}(2015)}]{Kneur:2015dda}%
\BibitemOpen
\bibfield  {author} {\bibinfo {author} {\bibfnamefont {J.-L.}\ \bibnamefont
		{Kneur}}\ and\ \bibinfo {author} {\bibfnamefont {A.}~\bibnamefont {Neveu}},\
}\href {\doibase 10.1103/PhysRevD.92.074027} {\bibfield  {journal} {\bibinfo
	{journal} {Phys. Rev.}\ }\textbf {\bibinfo {volume} {D92}},\ \bibinfo {pages}
{074027} (\bibinfo {year} {2015})}\BibitemShut {NoStop}%
\bibitem [{\citenamefont {Terning}(1991)}]{Terning:1991yt}%
\BibitemOpen
\bibfield  {author} {\bibinfo {author} {\bibfnamefont {J.}~\bibnamefont
		{Terning}},\ }\href@noop {} {\bibfield  {journal} {\bibinfo  {journal} {Phys.
			Rev.}\ }\textbf {\bibinfo {volume} {D44}},\ \bibinfo {pages} {887} (\bibinfo
	{year} {1991})}\BibitemShut {NoStop}%
\bibitem [{\citenamefont {Dorokhov}\ \emph {et~al.}(2015)\citenamefont
	{Dorokhov}, \citenamefont {Radzhabov},\ and\ \citenamefont
	{Zhevlakov}}]{Dorokhov:2015psa}%
\BibitemOpen
\bibfield  {author} {\bibinfo {author} {\bibfnamefont {A.~E.}\ \bibnamefont
		{Dorokhov}}, \bibinfo {author} {\bibfnamefont {A.~E.}\ \bibnamefont
		{Radzhabov}}, \ and\ \bibinfo {author} {\bibfnamefont {A.}~\bibnamefont
		{Zhevlakov}},\ }\href@noop {} {\bibfield  {journal} {\bibinfo  {journal}
		{Eur. Phys. J.}\ }\textbf {\bibinfo {volume} {C75}},\ \bibinfo {pages} {417}
	(\bibinfo {year} {2015})}\BibitemShut {NoStop}%
\bibitem [{\citenamefont {Weinberg}(1963)}]{Weinberg:1962hj}%
\BibitemOpen
\bibfield  {author} {\bibinfo {author} {\bibfnamefont {S.}~\bibnamefont
		{Weinberg}},\ }\href {\doibase 10.1103/PhysRev.130.776} {\bibfield  {journal}
	{\bibinfo  {journal} {Phys. Rev.}\ }\textbf {\bibinfo {volume} {130}},\
	\bibinfo {pages} {776} (\bibinfo {year} {1963})}\BibitemShut {NoStop}%
\bibitem [{\citenamefont {Salam}(1962)}]{Salam:1962ap}%
\BibitemOpen
\bibfield  {author} {\bibinfo {author} {\bibfnamefont {A.}~\bibnamefont
		{Salam}},\ }\href {\doibase 10.1007/BF02733330} {\bibfield  {journal}
	{\bibinfo  {journal} {Nuovo Cim.}\ }\textbf {\bibinfo {volume} {25}},\
	\bibinfo {pages} {224} (\bibinfo {year} {1962})}\BibitemShut {NoStop}%
\bibitem [{\citenamefont {Dorokhov}\ \emph {et~al.}(2012)\citenamefont
	{Dorokhov}, \citenamefont {Radzhabov},\ and\ \citenamefont
	{Zhevlakov}}]{Dorokhov:2012qa}%
\BibitemOpen
\bibfield  {author} {\bibinfo {author} {\bibfnamefont {A.~E.}\ \bibnamefont
		{Dorokhov}}, \bibinfo {author} {\bibfnamefont {A.~E.}\ \bibnamefont
		{Radzhabov}}, \ and\ \bibinfo {author} {\bibfnamefont {A.~S.}\ \bibnamefont
		{Zhevlakov}},\ }\href {\doibase 10.1140/epjc/s10052-012-2227-3} {\bibfield
	{journal} {\bibinfo  {journal} {Eur. Phys. J.}\ }\textbf {\bibinfo {volume}
		{C72}},\ \bibinfo {pages} {2227} (\bibinfo {year} {2012})}\BibitemShut
{NoStop}%
\bibitem [{\citenamefont {Dorokhov}\ and\ \citenamefont
	{Broniowski}(2003)}]{Dorokhov:2003kf}%
\BibitemOpen
\bibfield  {author} {\bibinfo {author} {\bibfnamefont {A.~E.}\ \bibnamefont
		{Dorokhov}}\ and\ \bibinfo {author} {\bibfnamefont {W.}~\bibnamefont
		{Broniowski}},\ }\href {\doibase 10.1140/epjc/s2003-01386-x} {\bibfield
	{journal} {\bibinfo  {journal} {Eur. Phys. J.}\ }\textbf {\bibinfo {volume}
		{C32}},\ \bibinfo {pages} {79} (\bibinfo {year} {2003})}\BibitemShut
{NoStop}%
\bibitem [{\citenamefont {Osipov}\ \emph {et~al.}(2007)\citenamefont {Osipov},
	\citenamefont {Radzhabov},\ and\ \citenamefont {Volkov}}]{Osipov:2007zz}%
\BibitemOpen
\bibfield  {author} {\bibinfo {author} {\bibfnamefont {A.}~\bibnamefont
		{Osipov}}, \bibinfo {author} {\bibfnamefont {A.}~\bibnamefont {Radzhabov}}, \
	and\ \bibinfo {author} {\bibfnamefont {M.}~\bibnamefont {Volkov}},\
}\href@noop {} {\bibfield  {journal} {\bibinfo  {journal} {Phys.Atom.Nucl.}\
}\textbf {\bibinfo {volume} {70}},\ \bibinfo {pages} {1931} (\bibinfo {year}
{2007})}\BibitemShut {NoStop}%
\bibitem [{\citenamefont {Hannah}(2001)}]{Hannah:2001ee}%
\BibitemOpen
\bibfield  {author} {\bibinfo {author} {\bibfnamefont {T.}~\bibnamefont
		{Hannah}},\ }\href@noop {} {\bibfield  {journal} {\bibinfo  {journal} {Nucl.
			Phys.}\ }\textbf {\bibinfo {volume} {B593}},\ \bibinfo {pages} {577}
	(\bibinfo {year} {2001})}\BibitemShut {NoStop}%
\bibitem [{\citenamefont {Truong}(2002)}]{Truong:2001en}%
\BibitemOpen
\bibfield  {author} {\bibinfo {author} {\bibfnamefont {T.~N.}\ \bibnamefont
		{Truong}},\ }\href@noop {} {\bibfield  {journal} {\bibinfo  {journal} {Phys.
			Rev.}\ }\textbf {\bibinfo {volume} {D65}},\ \bibinfo {pages} {056004}
	(\bibinfo {year} {2002})}\BibitemShut {NoStop}%
\bibitem [{\citenamefont {Hoferichter}\ \emph {et~al.}(2012)\citenamefont
	{Hoferichter}, \citenamefont {Kubis},\ and\ \citenamefont
	{Sakkas}}]{Hoferichter:2012pm}%
\BibitemOpen
\bibfield  {author} {\bibinfo {author} {\bibfnamefont {M.}~\bibnamefont
		{Hoferichter}}, \bibinfo {author} {\bibfnamefont {B.}~\bibnamefont {Kubis}},
	\ and\ \bibinfo {author} {\bibfnamefont {D.}~\bibnamefont {Sakkas}},\ }\href
{\doibase 10.1103/PhysRevD.86.116009} {\bibfield  {journal} {\bibinfo
		{journal} {Phys. Rev.}\ }\textbf {\bibinfo {volume} {D86}},\ \bibinfo {pages}
	{116009} (\bibinfo {year} {2012})}\BibitemShut {NoStop}%
\bibitem [{\citenamefont {Cotanch}\ and\ \citenamefont
	{Maris}(2003)}]{Cotanch:2003xv}%
\BibitemOpen
\bibfield  {author} {\bibinfo {author} {\bibfnamefont {S.~R.}\ \bibnamefont
		{Cotanch}}\ and\ \bibinfo {author} {\bibfnamefont {P.}~\bibnamefont
		{Maris}},\ }\href@noop {} {\bibfield  {journal} {\bibinfo  {journal} {Phys.
			Rev.}\ }\textbf {\bibinfo {volume} {D68}},\ \bibinfo {pages} {036006}
	(\bibinfo {year} {2003})}\BibitemShut {NoStop}%
\bibitem [{\citenamefont {Briceno}\ \emph {et~al.}(2015)\citenamefont
	{Briceno}, \citenamefont {Dudek}, \citenamefont {Edwards}, \citenamefont
	{Shultz}, \citenamefont {Thomas},\ and\ \citenamefont
	{Wilson}}]{Briceno:2015dca}%
\BibitemOpen
\bibfield  {author} {\bibinfo {author} {\bibfnamefont {R.~A.}\ \bibnamefont
		{Briceno}}, \bibinfo {author} {\bibfnamefont {J.~J.}\ \bibnamefont {Dudek}},
	\bibinfo {author} {\bibfnamefont {R.~G.}\ \bibnamefont {Edwards}}, \bibinfo
	{author} {\bibfnamefont {C.~J.}\ \bibnamefont {Shultz}}, \bibinfo {author}
	{\bibfnamefont {C.~E.}\ \bibnamefont {Thomas}}, \ and\ \bibinfo {author}
	{\bibfnamefont {D.~J.}\ \bibnamefont {Wilson}},\ }\href@noop {} {\bibfield
	{journal} {\bibinfo  {journal} {Phys. Rev. Lett.}\ }\textbf {\bibinfo
		{volume} {115}},\ \bibinfo {pages} {242001} (\bibinfo {year}
	{2015})}\BibitemShut {NoStop}%
\bibitem [{\citenamefont {Briceño}\ \emph {et~al.}(2016)\citenamefont
	{Briceño}, \citenamefont {Dudek}, \citenamefont {Edwards}, \citenamefont
	{Shultz}, \citenamefont {Thomas},\ and\ \citenamefont
	{Wilson}}]{Briceno:2016kkp}%
\BibitemOpen
\bibfield  {author} {\bibinfo {author} {\bibfnamefont {R.~A.}\ \bibnamefont
		{Briceño}}, \bibinfo {author} {\bibfnamefont {J.~J.}\ \bibnamefont {Dudek}},
	\bibinfo {author} {\bibfnamefont {R.~G.}\ \bibnamefont {Edwards}}, \bibinfo
	{author} {\bibfnamefont {C.~J.}\ \bibnamefont {Shultz}}, \bibinfo {author}
	{\bibfnamefont {C.~E.}\ \bibnamefont {Thomas}}, \ and\ \bibinfo {author}
	{\bibfnamefont {D.~J.}\ \bibnamefont {Wilson}},\ }\href@noop {} {\bibfield
	{journal} {\bibinfo  {journal} {Phys. Rev.}\ }\textbf {\bibinfo {volume}
		{D93}},\ \bibinfo {pages} {114508} (\bibinfo {year} {2016})}\BibitemShut
{NoStop}%
\bibitem [{\citenamefont {Al~Ghoul}\ \emph {et~al.}(2016)\citenamefont
	{Al~Ghoul} \emph {et~al.}}]{Ghoul:2015ifw}%
\BibitemOpen
\bibfield  {author} {\bibinfo {author} {\bibfnamefont {H.}~\bibnamefont
		{Al~Ghoul}} \emph {et~al.} (\bibinfo {collaboration} {GlueX}),\ }\bibfield
{booktitle} {\emph {\bibinfo {booktitle} {{Proceedings, 16th International
				Conference on Hadron Spectroscopy (Hadron 2015): Newport News, Virginia, USA,
				September 13-18, 2015}}},\ }\href@noop {} {\bibfield  {journal} {\bibinfo
		{journal} {AIP Conf. Proc.}\ }\textbf {\bibinfo {volume} {1735}},\ \bibinfo
	{pages} {020001} (\bibinfo {year} {2016})}\BibitemShut {NoStop}%
\end{thebibliography}

%

\end{document}